\documentclass[twocolumn,showpacs,preprintnumbers,amsmath,amssymb]{revtex4} 

 \usepackage{graphicx}
 \usepackage{dcolumn}
 \usepackage{bm}

\begin{document}

 \title{Reply to ``Comment on Jamming at zero temperature and zero applied
stress: The epitome of disorder''} 
 \author{Corey S. O'Hern} 
 \affiliation{ Department of Mechanical Engineering, 
 Yale University, New Haven, CT  06520-8284.} 
 \author{Leonardo E. Silbert} 
 \affiliation{ Department of Chemistry and Biochemistry, UCLA, Los 
 Angeles, CA  90095-1569 and The James Franck Institute, The University   
 of Chicago, 
 Chicago, IL 60637.} 
 \author{Andrea J. Liu} 
 \affiliation{Department of Chemistry and Biochemistry, UCLA, Los 
 Angeles, CA  90095-1569} 
 \author{Sidney R. Nagel} 
 \affiliation{James Franck Institute, The University of Chicago, 
 Chicago, IL 60637} 
 \date{\today} 

\begin{abstract} 
We answer the questions raised by Donev, Torquato, Stillinger, and
Connelly in their Comment on ``Jamming at zero temperature and zero
applied stress: The epitome of disorder.''  We emphasize that we
follow a fundamentally different approach than they have done to
reinterpret random close packing in terms of the ``maximally random
jammed'' framework.  We define the ``maximally random jammed packing
fraction'' to be where the largest number of initial states, chosen
completely randomly, have relaxed final states at the jamming
threshold in the thermodynamic limit.  Thus, we focus on an ensemble
of states at the jamming threshold, while DTSC are interested in
determining the amount of order and degree of jamming for a particular
configuration.  We also argue that soft-particle systems are as
``clean'' as those using hard spheres for studying jammed packings and
point out the benefits of using soft potentials.
\end{abstract} 


 \maketitle 

 \section{Overview:  What is the difference between the two approaches?} 

 The meaning of random close-packing is fraught with ambiguities.  A 
 given amorphous packing can be made slightly more dense by introducing 
 small amounts of crystallinity; thus, the concept of ``randomness" and 
 the concept of ``close-packing" would appear to be at odds with one 
 another. 

 Torquato and coworkers\cite{torquato} have pioneered a 
 re-examination of random close-packing (RCP) in terms of the notion of a 
 ``maximally random jammed" (MRJ) state, with specific definitions of 
 ``maximally random" and ``jammed." 
 The point of view espoused by Torquato, et al.~\cite{torquato} and by 
 Donev, Torquato, Stillinger and Connelly \cite{donev} 
 (DTSC) is fundamentally different from the one we have adopted   
 ~\cite{prl,jlong}. 
 They seek to identify, for a specific, finite 
 configuration of hard spheres, the degree to which that configuration 
 can be considered maximally random and jammed.  They introduce three 
 different categories of jammed states and employ a series of order 
 parameters to measure the magnitude of different possible forms of 
 order.  A given configuration is maximally random if all of these order 
 parameters are minimized with respect to variations of the particle   
 positions and 
 lattice vectors of the periodic cell.  Thus, 
 their emphasis is on finding the amount of order and degree of jamming 
 in any given configuration. 

 By contrast, our point of view does not seek to identify the degree of 
 order of any specific configuration.  We are exclusively interested in 
 defining an ensemble of states that are at the threshold of jamming. 
 The results we have quoted were obtained by extrapolation to the 
 thermodynamic limit not for infinitely hard spheres, but for soft 
 particles that can overlap.  In our studies, we have considered a 
 configuration to be jammed if both the bulk and shear moduli are 
 nonzero.  One remarkable finding was that both of these moduli have 
 their thresholds at the same packing density for all configurations that 
 we studied.  In addition, we examined the spectrum of vibrational modes 
 and found that above the jamming threshold, all modes have nonzero 
 frequency~\cite{remark}.  In our case, the ``maximally random" density   
 is defined 
 in terms of an ensemble of configurations constructed as follows. 
 Because we use soft 
 particles, we can initially place $N$ particles of volume $v$ at random 
 within a box of size $L^d$, where $L$ is the box length and $d$ is the 
 dimensionality of space.  (This corresponds to infinite temperature and 
 cannot be done with hard 
 spheres, which are never allowed to overlap.) Using a conjugate gradient 
 or steepest descent algorithm, we relax the initial configuration at 
 fixed packing fraction $\phi \equiv Nv/L^d$ to its nearest energy 
 minimum; this defines the final state.  This relaxation depends on the 
 inter-particle potential and not on any particle dynamics.  It is   
 therefore 
 a property of the potential energy landscape.  For a given number of 
 particles $N$, we define the ``maximally random jammed {\it packing   
 fraction}" 
 to be where the highest number of initial states have final states at   
 the 
 jamming threshold.  As we take the thermodynamic limit $N \rightarrow 
 \infty$, we find that the width of the distribution of jamming 
 thresholds approaches zero; this indicates that virtually all of the 
 configurations, {\it which were sampled randomly}, jam at the same   
 packing 
 fraction. 
 The value of this packing fraction corresponds to the 
 number commonly associated with 
 random close-packing. 
 We verified that this distribution of jamming thresholds does not depend 
 on the potentials we chose. 
 Thus, while any given configuration can be jammed or not jammed, the 
 ``maximally random" density can be defined only by considering an   
 ensemble 
 of configurations. 

 The two approaches, that of DTSC and our own, are similar in that they 
 re-interpret ``random close-packing" in terms of the ``maximally random 
 jammed" terminology \cite{torquato}.    We will argue that the two   
 approaches are 
 equally valid. 
 In Section II, we respond to specific comments of DTSC.  However, as we   
 discuss in 
 Section III, the question is not which of these 
 approaches is more valid, but which is more useful. 

 \section{Response to specific comments} 

 \subsection{What is ``Jammed"?} 
 DTSC argue that we do not distinguish between the three different levels 
 of jamming defined in Ref. ~\cite{ts}, namely ``local," ``collective" and 
 ``strict" jamming.  It is indeed true that we are not interested in 
 ``local" jamming, the least restrictive of their definitions, in which 
 groups of particles are free to move.  Instead, we are 
 interested in systems where the bulk and shear moduli are nonzero. By 
 also ensuring that all the vibrational modes have positive frequency (in 
 other words, that the dynamical matrix is positive definite so that   
 configurations that 
 are only locally jammed are excluded), we not only 
 guarantee that the moduli are nonzero but also that the system is   
 isostatic at the jamming 
 threshold.  In an isostatic system, the elastic properties are   
 independent of inter-particle 
 potential and thus dependent only on the geometry of the configuration. 
 Thus, the soft-particle system is as ``clean" as hard-sphere systems for 
 studying the purely geometrical properties of the physical point J. 

 DTSC demonstrate in their Comment that our definition of ``jamming" is 
 closely related to their definition of ``collective jamming." 
 As we have said in the Overview, we are interested in the 
 thermodynamic limit, when the number of particles in our system 
 approaches infinity.  In such a limit, boundary conditions no longer 
 affect whether or not a system is jammed.  Thus, the distinction between 
 their definitions of ``collective" and ``strict" jamming disappears. 

 \subsection{What is ``Random"?} 

 In our framework, we have concentrated on creating a completely random   
 set of 
 configurations for the initial state.  In our view, this is where   
 ``randomness" 
 enters the problem.  We thus find the fraction of all phase space that   
 is 
 funneled down (upon relaxation) into jammed configurations (i.e. the   
 fraction of 
 phase space that has inherent structures that are jammed).  This   
 sampling 
 can easily be done with soft particles, but is impossible with hard   
 spheres. 
 For our systems, this provides a consistent and well-defined ensemble   
 with which 
 to work. 

 We agree with DTSC 
 that we would also like to know the distribution of all 
 possible final states at zero energy that are at the jamming threshold.   
   If that knowledge 
 were available, then it would be possible to define ``maximally random   
 jammed" 
 with reference only to the final states.  Unfortunately, an algorithm   
 to find such a distribution 
 is unavailable.  Such an approach would be complementary to ours but   
 would not supplant it. 

 We note that all of our {\it final} states at zero energy (i.e. states   
 at or below the jamming threshold) 
 are allowed hard-sphere states.  Presumably 
 this is why DTSC are particularly interested in the randomness of final   
 configurations 
 as opposed to initial ones.  We can take the limit of using harder and   
 harder potentials to see if 
 any of our initial-state distributions change on approaching the   
 hard-sphere limit.  We used 
 $V(r) = \epsilon \alpha^{-1} (1-r/\sigma)^\alpha$ for $r<\sigma$, $V(r)=0$   
 for $r \ge \sigma$, 
 where $\sigma$ is the particle diameter~\cite{remark2}.  We find that   
 the distributions 
 are indistinguishable for three different 
 values of $\alpha$, namely 5/2, 2, and 3/2. 
 It is for this reason that we believe that our results 
 are relevant to studies of hard-sphere systems.  Of course, since our   
 data are numerical and 
 because there are different ways of taking the hard-sphere limit,   
 including qualitatively different 
 kinds of potentials, one can always worry that the results might change   
 as one 
 approaches the hard-sphere limit more closely. 

 In order to define the 
 ``maximally random jammed" density, we focus only on distributions of   
 configurations.  From our 
 point of view, randomness does not describe a particular configuration,   
 but rather the 
 ensemble of initial states.  Contrary to the assertion of DTSC, we are   
 not proposing a unique 
 definition of order for the ensemble.  Following their example of a   
 jammed but diluted FCC 
 lattice packing, this would be an allowed but {\it highly} improbable   
 state in our distribution. 

 The example of a two-dimensional monodisperse disk packing is more
 problematic \cite{donev2}.  It is unclear for such a situation
 whether the phrases ``maximally-random jammed" or ``random
 close-packed" are appropriate.  Indeed if one {\it only} had such a
 system one would never have come up with the idea of random
 close-packing.  However, our definition in terms of the ensemble
 would still yield a well-defined MRJ density.  The identification of
 various types of order in a given packing is a deep and interesting
 question.  However, it is unclear whether one wants to conflate that
 issue with a definition of an RCP or MRJ density.  It depends on what
 one wants to learn from the definition.

 \subsection{Universal algorithms} 

 Contrary to the assertion of DTSC, we are not claiming to explore the   
 space of all jammed 
 configurations in an unbiased manner.  Rather, we are exploring the   
 space of all initial 
 states (at infinite temperature, $T=\infty$) in an unbiased manner.   
 DTSC see no difference 
 between starting at $T=\infty$ or any other temperature; the advantage   
 of starting at $T=\infty$ is 
 that one can at least sample initial states completely randomly. 

We are glad that DTSC have pointed out something that may have been 
confusing in our paper.  Because we have used two different protocols 
to determine different types of results in our paper, they believe that 
we have mixed them up in determining the distribution of jamming 
thresholds, $P_j(\phi)$.  This is not so.  In determining $P_j(\phi)$ 
(Fig. 6 of Ref.~\cite{jlong}) we used the first protocol where we never 
varied the volume of our system during the relaxation process.  We have 
said this explicitly in Section II.C of our paper. 
To find the {\it distribution} of jamming thresholds, it is not 
necessary to determine the value of $\phi_c$ for any given {\it 
configuration}.  Instead we only need to know the fraction of states, 
$f_j(\phi)$  that are jammed at any value of $\phi$.  $P_j(\phi)$ is 
then the derivative of $f_j(\phi)$ with respect to $\phi$.  In order to 
find $f_j(\phi)$ we need only determine if a state, produced by 
relaxation at a {\it fixed} packing fraction, is jammed or unjammed.   
This is, as we said earlier, easy to do since it only involves 
calculating whether the final configuration has a positive-definite   
pressure, shear modulus or dynamical matrix.  This knowledge, in and of 
itself, is sufficient to determine whether the state is jammed.   
Whether a configuration is jammed or not does not require a knowledge 
of the precise value of $\phi_c$ for that state. 
We should point out that we could have obtained our results for the 
coincidence of the pressure and the shear modulus approaching zero at 
the same value of $\phi$ without ever using a compression or 
decompression run (the second protocol) but simply by plotting 
parametrically the pressure versus shear modulus for all states 
obtained by relaxation at fixed packing fraction (the first protocol). 

DTSC objected that obtaining the distribution of jamming thresholds by
looking at the fraction of jammed states is unphysical for hard
spheres.  Note that states that are at or below the jamming threshold
are allowed hard-sphere packings.  The fraction of such unjammed
configurations $f_u(\phi)$, relevant to hard-sphere packings, is
simply $1-f_j(\phi)$.  Thus, the same distribution of jamming
thresholds could have been obtained just as easily from the fraction
of unjammed, hard-sphere, states.

 \section{Why our approach is useful} 

 The question of which approach is more useful depends on what one wants   
 to investigate.  Perhaps the 
 most important advantage of studying soft particles is that we can   
 study properties of packings 
 both above and below RCP or MRJ density.  This allows a more complete   
 picture of the properties of 
 Point J.  For example, the divergence of the pair correlation function   
 $g(r)$ near $r=\sigma$, 
 $g(r) \sim (r-\sigma)^{-1/2}$~\cite{leo,prl,jlong} was completely missed   
 by studies of hard spheres, 
 but was uncovered by using softer potentials.  We have also found that   
 the properties 
 of a jammed configuration depend solely on $\phi-\phi_c$, that is, the   
 distance from the jamming 
 threshold (these studies were the only ones in which we allowed the   
 density to vary, using 
 the second procedure described by DTSC).  Because we have shown that,   
 at threshold, our states 
 are isostatic, the mechanical properties of our system at $\phi_c$   
 (which approaches the RCP density 
 in the infinite-system size limit) do not depend on the potential   
 chosen but depend only on 
 the geometry of the configuration. 
 Thus, the soft-particle system is as ``clean" as hard-sphere systems for 
 studying the purely geometrical properties of Point  J. 

 Grant support from NSF-DMR-0087349 (LES,AJL), DE-FG02-03ER46087   
 (LES,AJL), 
 NSF-DMR-0089081 (LES,SRN) and DE-FG02-03ER46088 (LES,SRN) is 
 gratefully acknowledged.


\begin{references}
\bibitem{remark2} We studied both monodisperse and bidisperse systems,   
 where $\sigma$ varies among 
 particle pairs and is the sum of the radii of the two particles.
\end{references}
 \end{document}